\documentclass[referee]{aa} 

\usepackage{graphicx}
\begin{document}

   \title{Galaxy Pairs in cosmological simulations: Effects of interactions
on star formation.}

   \author{M. Josefa Perez 
          \inst{1}\inst{2}\inst{3},
           Patricia B. Tissera
          \inst{1}\inst{3},
           Diego G. Lambas
          \inst{1}\inst{4},
          \and
           Cecilia Scannapieco
          \inst{1}\inst{3}
          }

   \offprints{M. Josefa Perez}

   \institute{Consejo Nacional de Investigaciones Cient\'{\i}ficas y T\'ecnicas, Argentina.\\
         \and 
              Facultad de Ciencias Astron\'omicas y Geof\'{\i}sicas, La Plata, Argentina.\\
              \email{jperez@fcaglp.unlp.edu.ar}
         \and 
              Instituto de Astronom\'{\i}a y F\'{\i}sica del Espacio, Argentina.\\
              \email{patricia@iafe.uba.ar}
         \and 
              Observatorio Astron\'omico de la Universidad Nacional de C\'ordoba, Argentina.\\
             }


   \abstract{
We carried out a statistical analysis of galaxy pairs in hydrodynamical 
$\Lambda$CDM simulations. We focused on the triggering of star formation by
interactions and analysed the enhancement of star formation activity in
terms of orbital parameters. By comparing to a suitable sample of simulated galaxies
 without a nearby companion, we find that  close 
 encounters ($r<30$ kpc $h^{-1}$)  may
effectively induce star formation.
However, our results suggest that the stability properties of systems and
the spatial proximity are both relevant factors in the process of triggering
star formation by tidal interactions.
In order to assess the effects of projection and spurious pairs in
observational samples, we  also constructed and analysed samples of
pairs of galaxies in the simulations obtained in projection. We found a
good agreement with observational results with a threshold at $r_{\rm p} \sim 25
$  kpc $h^{-1}$ for interactions to {\bf statistically} enhance star formation activity.
 For pairs within $r_{\rm p} < 100
$  kpc $h^{-1}$, we estimated a $\sim 27 \%$ contamination by spurious pairs, 
reduced to $\sim 19 \%$ for close
systems.   We also found  that spurious pairs affect more strongly high density regions with $17\%$ of spurious
pairs detected for low density regions compared to $33\%$ found in high density ones. 
Also, we  analysed the dependence of star formation on environment by
defining the usual projected density parameter for both  pairs and isolated
galaxies in the simulations. We find the expected star formation-local density
relation for  both galaxies in pairs and without a close companion, with a stronger  density
dependence for close pairs which  suggests a  relevant role for interactions  in
driving this relation.

   }
            
   \authorrunning{M. Josefa Perez et al}
   
   \titlerunning{Galaxy Pairs in cosmological simulations}
   
   \maketitle
%

\section{Introduction}

If the Universe evolves according to the 
current cosmological paradigm which postulates
the hierarchical  growth of the structure, mergers and interactions
are common events in the life of  galaxies.
The effects of these violent processes are highly complex to study
owing to their non-linearity and  their possible dependence  on redshift.
Toomre et al. (1976) built up the first numerical simulations of 
galaxy interactions showing that many observed characteristics
of galaxies, such as  morphology,  could be explained by these processes.
These authors showed that galactic discs are unstable under tidal interactions
which could modify the mass distribution,  
converting spiral and irregular galaxies into bulge ellipticals and SOs.
 More recent numerical simulations of pre-prepared mergers showed that 
interactions between axisymmetrical systems without bulges or with  
small ones might induce  gas inflows to the central region of the systems,
triggering  starburst episodes (e.g. Barnes \& Hernquist 1996;
Mihos \& Hernquist 1996). Using cosmological hydrodynamical simulations, 
Tissera et al. (2002) studied the effects of mergers in the star formation
history of galactic objects in hierarchical 
clustering scenarios, finding similar results.
These results indicated that, during some merger events, 
gaseous discs could experience two starbursts depending
on the characteristics of the potential wells of the systems.
The first starburst is triggered during the 
orbital decay phase by  gas inflows driven as the
satellites approach each other, while the second one 
is produced  when the  two baryonic clumps collide, in agreement with results from
pre-prepared simulations.
These authors  also showed that the effects of 
interactions on the star formation activity
were different at different stages of evolution of the systems, 
being more efficient  in early phases when 
their potential wells were shallower. 

In the local Universe, observations showed that mergers and interactions
 affect the star formation (SF) activity  
in galaxies (e.g. Larson \& Tinsley 1978;
Donzelli \& Pastoriza 1997). Observations of high redshift systems found 
that the merger rate and the star formation activity of galaxies
increase with redshift, suggesting a change in the impact of interactions on
the SF process as galaxies evolve (Le F\`evre et al. 2000; Patton et al 2002).
Barton et al. (2000)
showed that interactions during close encounters could be
correlated with enhancements of the star formation
activity. These results were confirmed by 
Lambas et al. (2003) and Alonso et al. (2004) who explored 
the dependence of star formation activity 
in galaxies with close companions on  relative proximity 
and  environment by constructing galaxy
pairs catalogs from the Two Degree Field Galaxy Redshift Survey 
(2dFGRS; Colles et al 2001).
These authors found that independently of environment, 
galaxy pairs exhibit an enhancement of the star formation 
activity with respect to galaxies without a close companion from approximately 
the same  projected distance and relative radial velocity 
thresholds: r$_{\rm p}\leq$ 25 kpc $h^{-1}$ and $\Delta V \leq$ 100 km s$^{-1}$.
This independence of the projected orbital thresholds 
suggests that  the nature of star formation activity
driven by galaxy interactions may be independent of environment,
although  the general  level of star formation activity tends to be  lower
in high density regions (see also Nikolic et al. 2004).

A main shortcoming of studying galaxy pairs in 
observed catalogs is the impossibility of estimating the 
3D separation between galaxies. There are two main 
sources of problems: the geometrical effects of projecting
and the formation of  spurious pairs owing to the projection.
 Although  Lambas et al. (2003) and Alonso et al. (2004)
 analysed one of the largest  available galaxy pairs samples,
these effects could introduce noise or distort the correlation signals. 
>From different numerical approaches, Mammon  (1986) 
and Alonso et al. (2004) found
that the effects of spurious pairs tend to be  larger 
in high density environments and
for larger relative projected separations.
In this paper, we  analyse hydrodynamical cosmological 
simulations and constructed  3D and 2D catalogs of galaxy pairs
 (hereafter GPs) with the aim at confronting 
the current cosmological paradigm with
observations and assessing the impact of the  
two mentioned projection effects on the results.
 The 3D-GPs were selected by proximity criterium while for the 
2D-GPs  we applied  both relative velocity 
and spatial separation cut-offs following the observational 
procedures described by    Lambas et al. (2003). 

This paper is structured as follows: Section 2 describes 
the main characteristics of the
numerical simulations and discusses the galaxy pairs selection.
Section 3 shows results  for the tridimensional galaxy pair catalog. 
In Section 4, we analyse  the projected galaxy sample and confront it with observations.
Finally, we summarize the findings in Section 5.


\section{Catalogs of Galaxy pairs in numerical simulations.}

In this Section we describe the main characteristics of the analysed
simulations and the procedure applied for the identification of
galactic systems and the pair definition.

\subsection{Numerical Experiments}

We discuss results for  cosmological simulations 
that include the gravitational and hydrodynamical
evolution of the matter, star formation, chemical 
enrichment and metal-dependent cooling.
We have performed a  $\Lambda$CDM simulation with 
the chemical GADGET-2 of Scannapieco et al. (2005).
We have also analysed a  $\Lambda =0$ experiment 
(hereafter SCDM) with the chemical hydrodynamical code of 
Mosconi et al. (2001)  to examine at what extent 
galaxy-galaxy interactions are affected
by the cosmological parameters. Table 1 summarizes 
the main simulation parameters.
The simulated volume is representative of a typical
field region of the Universe but it is smaller than 
that covered by current galaxy surveys.
Hence, comparison with  observations are only intended to
constrain global trends.

In both simulations the star formation algorithms are  based on the 
Schmidt law (Navarro \& White 1994; 
Tissera  2000). However, the chemical GADGET-2  
transforms gas into stars in a stochastic way, avoiding the use
of hybrid particles (i.e. particles representing 
gas and stars at the same time).
Both chemical codes describe the enrichment 
of the interstellar medium  by SNII and SNIa 
 (see for details  Mosconi et al. (2001) and Scannapieco et al. (2005)),
following the production of the same chemical elements.
The simulation SDCM has been used by Tissera et al. (2001, 2002) and
Cora et al. (2003) to study Damped Lyman Alpha Systems.

The initial particle masses  of the SCDM and $\Lambda$CDM 
simulations are different. 
The SCDM run has equal mass particles for
the baryons and dark matter of $M_{\rm p}= 1.3\times 10^{8} M_{\odot} h^{-1}$, 
while the  $\Lambda$CDM run has 
initially  $M_{\rm gas}= 2 \times 10^{7} M_{\odot} h^{-1}$ 
and $M_{\rm DM}=2 \times 10^{8} M_{\odot} h^{-1}$.
Hence, a comparison of the results from these two 
different cosmological simulations will also help 
to probe the robustness of our conclusions against
numerical resolution and codes. 

\begin{table*}
\begin{minipage}[t]{\columnwidth}
\caption{Summary of the main parameters of the  cosmological numerical simulations.}
\label{esp}
\centering
\renewcommand{\footnoterule}{}  
\centering    
\begin{tabular}{ c c c c c c c cccccccccc}
\hline
 Cosmology& $\Omega_{\rm m}$\footnote{$\Omega_{\rm m}$: Mass density parameter} & $\Omega_{\Lambda}$\footnote{$\Omega_{\Lambda}$: Vacuum energy density parameter} & $\Omega_{\rm b}$\footnote{$\Omega_{\rm b}$: Baryonic density parameter} & ${\rm H_0}$\footnote{$H_0$: Hubble constant}  &$\sigma_{8}$\footnote{$\sigma_{8}$: RMS density fluctuations at 8 Mpc $h^{-1}$}&$N_{\rm g}$\footnote{$N_{\rm g}$: Initial number of gas particles}&$N_{\rm d}\footnote{$N_{\rm d}$: Total number of dark matter particles}$&$L$\footnote{$L$: box size of the simulated volume Mpc $h^{-1}$} \\
\hline
 $\Lambda$CDM & 0.30  & 0.70 & 0.04 & 70 & 0.90 & $80^{3}$ &  $80^{3}$ &10&\\
 SCDM &  1.00 & 0.00 & 0.10 & 50 & 0.62 & 26214 & 235930&5&\\
\hline
\end{tabular}
\end{minipage}
\end{table*}

\subsection{Galaxy pair identification.}

The identification of  galaxy-like objects (GLOs) 
in the simulations was carried out according to the
following steps. First, we identified the global structures using the 
 percolation method  {\it friends-of-friends} (fof; Davis et al. 1985)
to select virialized  haloes.
This method allows us to  identify the 
gravitational bounded concentrations but not the substructures.
 In particular, when  two structures collide, their dark matter haloes will also do so.
Depending on the
particular dynamical characteristics of the encounter,
 the baryonic substructures
that can be associated with  galaxies  may be already sharing a common dark matter halo  at the
time we are looking at them, or not.
However, in this work, we are only
interested in the baryonic substructures within the dark matter haloes.
For this reason, we designed a  detailed procedure to individualize baryonic  satellites 
within  a region of   $0.5 \ {\rm  Mpc} \ h^{-1}$  
centered at each virialized system by
fine-tuning the  {\it linking length} parameter of the fof algorithm.
This  fine-tuning  was not possible 
to carry out automatically requiring a close check of 
the substructure selection in each region 
in order to prevent the inclusion of loose agglomerations.

By considering that the radius which encloses $83\%$ of the luminous mass of an exponential
disc corresponds to the isophote of 25 ${\rm mag \ arsec^{-2}}$ and by assuming that the mass-to-light
ratio is independent of radius, we define the optical radius as the one that encloses $83\%$ of the baryonic mass
of the system.
The GLOs analysed at two optical radius will be, hereafter,
 referred as  simulated galaxies.
After the  identification of the simulated galaxies, we analysed
their astrophysical  properties and 
their star formation activity within two optical radius.
We only take into account those systems with stellar masses
larger than   $8 \times 10^{8}\,{\rm  M_{\odot}}\, h^{-1}$ within two optical radius to
minimize numerical resolution problems. The final simulated galaxy sample is made
of 364 systems. 
For each simulated galaxy, we estimated the stellar 
birthrate parameter, $b=SFR/<SFR>$, defined 
as the present level of star formation  activity 
of a galaxy normalized to its mean past SF rate, $<SFR>$. Thus, systems
undergoing strong SF activity have $b>1$. This parameter
has been found to correlate with morphology
(Kennicutt 1998) in the sense that late-type 
spirals and starbursts have larger $b$ values.
Note that we  applied the physical concept behind
the birth rate parameter to define our simulated $b$ parameter.
Observations have to resort to galaxy formation models to estimate
the $<SFR>$ (Carter et al. 2002).
In the case of the simulations, we constructed the star formation
history of each GLO by estimating the SFR in each time step of
integration and then smoothing out the distributions with a
$10^7$ yr filter. This filter erases  numerical    
noise without deleting the main features.
Then, the simulated $b$ parameter is defined as a ratio
between the SFR at $z=0$ and the mean past star formation rate, $<SFR>$.

>From  the distribution of simulated galaxies 
we  built up the 3D catalog of galaxy pairs by applying
a proximity criteria which was determined 
by estimating the mean birth rate parameter in distance  bins.
This distribution showed a sharp increase 
of the star formation activity for $r  < 100\ {\rm kpc} \ h^{-1}$.
We take this distance threshold, $ r_{\rm cut}=100 \ {\rm kpc} \ h^{-1}$, 
to define galaxy pairs from the 3D distribution. Then, the so-called   
3D-GP catalog is made up of 88 galaxies in pairs. 
Systems with $ r_{\rm cut}> 100 \ {\rm kpc} \ h^{-1}$ 
have a mean star formation activity similar to the average of the simulated box.
In order to unveil the effects of having 
a close companion on the star formation activity,
we defined a control sample following the procedure  
applied by Lambas et al. (2003) to the
analysis of galaxy pairs in the 2dfGRS. 
This control sample   
is then  constructed by selecting those simulated galaxies which
do not have a companion within the distance threshold $ r_{\rm cut}$.

For the purpose of comparing our results with observations, 
a 2D simulated-galaxy pair catalog (hereafter 2D-GP catalog)
 was also constructed by projecting the total 
3D simulated-galaxy distribution onto random directions. 
We considered three different random observers which 
yield a total projected sample of 2184 simulated galaxies.
Then, the 2D  pairs were selected according 
to the observational criteria of Lambas et al. (2003).
Thus, the adopted thresholds in relative projected separation 
and radial velocity 
are: $r_{\rm p}=$ 100 kpc $h^{-1}$ and $\Delta cz=$ 350 km s$^{-1}$
which produce a sample of 677 simulated galaxies in pairs.
 Similarly, we define a close pair sample by adopting the observational criteria
determined by Lambas et al. (2003): $r_{\rm p}=$ 25 kpc $h^{-1}$ and $\Delta cz=$ 100 km s$^{-1}$,
obtaining a subsample of 194 close pairs in projection. 
The corresponding control sample was constructed 
from the projected simulated distribution  
by requiring its members  not to have
a companion system within these  thresholds.

\section{Analysis of the 3D Galaxy Pair Catalog}

In this Section we discuss the star formation activity 
as a function of orbital parameters in the 3D-GP catalog.
For this purpose we define the parameter 
$\beta= b / \langle b\rangle$, where $\langle b\rangle$ 
is the mean birthrate parameter of the control sample ($\langle b\rangle =0.7)$. 
Hence, $\beta$ provides a direct indication of the excess of
star formation in galaxy pairs with respect to the mean star formation
in galaxies without a close companion.

\begin{figure}
\centering
\includegraphics[width=7cm,height=5.5cm]{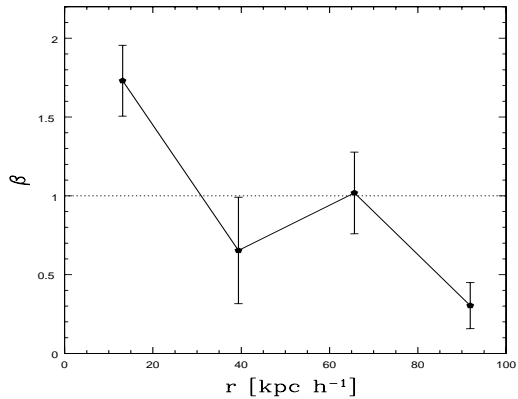}
\caption{Mean star formation excess parameter 
$\beta= b / \langle b\rangle$  (where $\langle b\rangle$ 
is the mean birthrate parameter for the control sample) 
as a function of the  tridimensional distance
to the closest neighbour in the $\Lambda$CDM experiment. Errors have been estimated 
by using the bootstrap resampling technique. } 
\label{gad3d-d}
\end{figure}

In Fig. \ref{gad3d-d}, we show  the mean $\beta$ parameter
as a function of the distance to the  closest neighbour. 
As we can see from this figure, there is a clear trend 
for an increase of the star formation
activity with proximity to a companion. 
The error bars have been estimated by applying the boostrap technique. On average, we have
20 galaxies per radius bin. 
Pairs closer than   the critical distance $r_{\rm c} = 30\pm 10 $ kpc $h^{-1}$ exhibit
an excess of star formation activity 
with respect to the mean SF of the control sample.
The error of $r_{\rm c}$  has been determined considering the $r$ values at the 
intersection of the corresponding 1-$\sigma$ error zone 
 with
the $\beta =1$ threshold. 
This  $r_{\rm c} = 30\pm 10$ kpc $h^{-1}$  
value will be taken as a 3D-distance threshold for
tidally triggered enhancement of the star formation activity. 

\begin{figure}
\centering
\includegraphics[width=7cm,height=5.5cm]{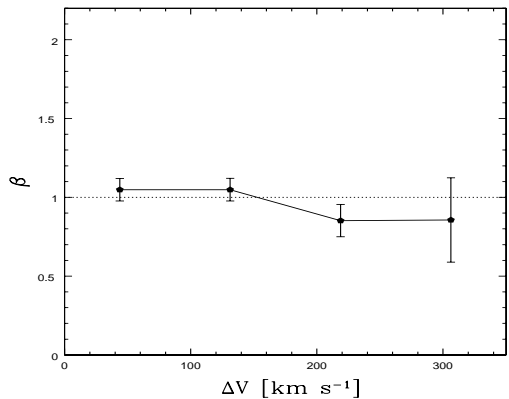}
\caption{Mean star formation excess parameter $\beta$   
as a function of the relative tridimensional velocity 
for galaxies pairs with $r<100$ kpc $h^{-1}$ in the $\Lambda$CDM experiment. 
Errors have been estimated by using the bootstrap resampling technique. } 
\label{gad3d-v}
\end{figure}

The analysis of the SF activity in simulated galaxy pairs
 as a function of  the relative tridimensional velocity 
shows  a flat  trend (Fig. \ref{gad3d-v}) with a very weak indication that
encounters with relative velocities lower than
$\Delta V_{\rm c} = 150\pm 25$ km s$^{-1}$  
may be statistically related with an  increase 
in the SF activity (Mihos 2004). The error for $\Delta V_{\rm c}$ has been estimated in a similar 
fashion to that of    $r_{\rm c}$.

We found that around $40\%$  of the simulated galaxy pairs, although 
having a companion within $r_{\rm c} = 30$ kpc $h^{-1}$,
do not show star formation enhancement. 
Previous works (e.g. Barnes \& Hernquist 1996; Tissera 2000; Tissera et al. 2002) have shown that,
although star formation can be induced during close-by interactions,
the triggering of gas inflows with subsequent strong star formation activity depends, at least,  on
the orbital parameters (Barnes \& Hernquist 1996), the gas reservoir and 
 the internal properties of the potential well (Tissera 2000).
Hence, strong star formation activity is not a {\bf necessarily the}  result of an interaction.
In this analysis we are taking into account systems with a variety of astrophysical and dynamical conditions and selecting them 
by using
only a proximity criterium. Then, we expect that a fraction of them will not exhibit strong star formation activity.
A similar situation might occur in observations.

If we separate galaxies in  pairs (with $r  < 100$ kpc $h^{-1}$), according to their star formation 
activity, in passive SF ($\beta < 1$) and
active SF ($\beta >1$) pairs, we get that a $66 \%$  of galaxies in pairs   do not have an excess of
star formation with respect to the mean of the control sample.
In order to understand the physical cause of active or passive star formation activity 
in galaxies with a close companion,  we analyse the properties of these systems.
This analysis is carried out on statistical basis since
we do not
have enough number of systems to segregate them according to all the parameters that might be relevant such as the different combination of orbital parameters.

\begin{figure}
\centering
\includegraphics[width=7cm,height=5.5cm]{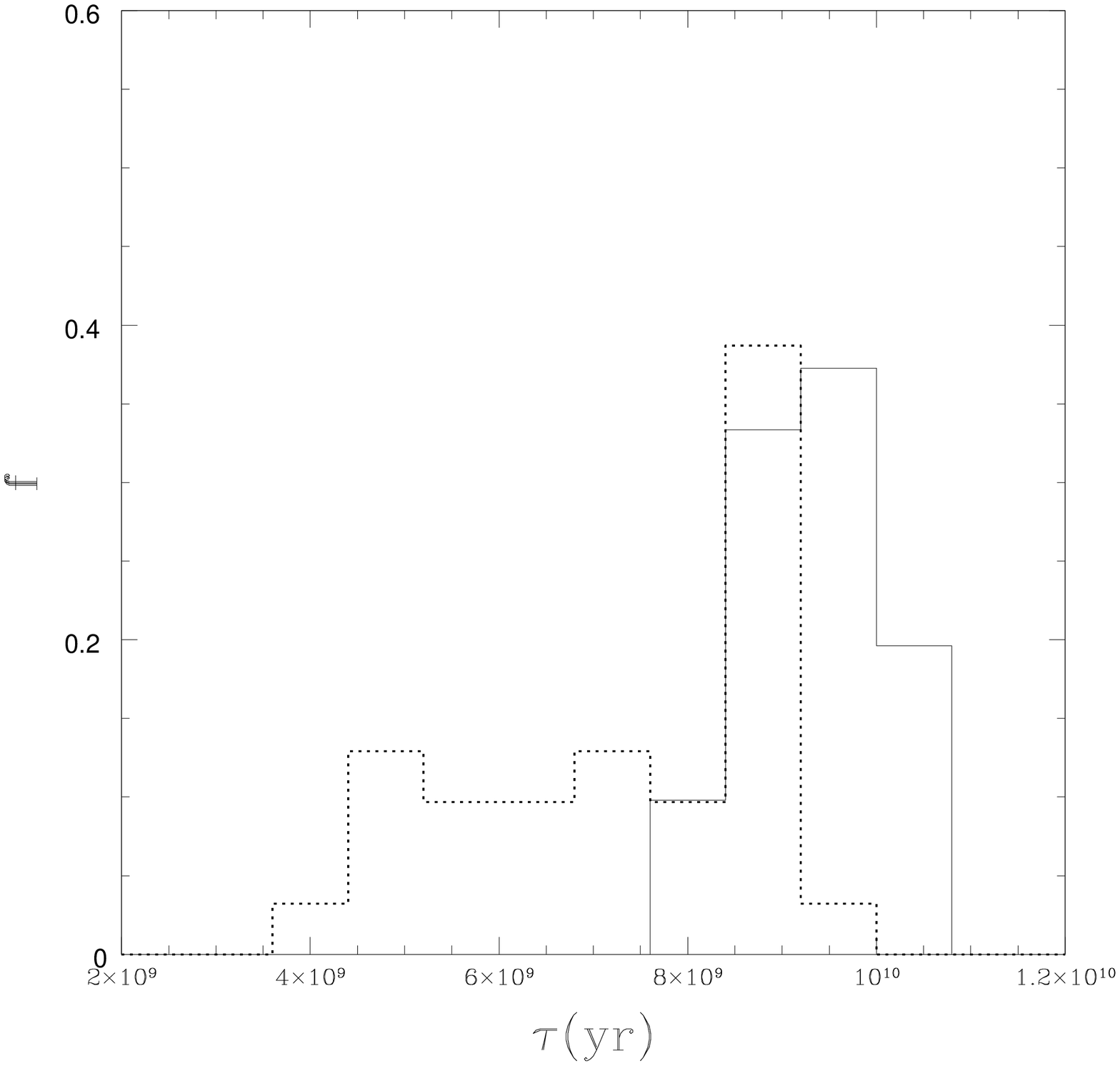}
\includegraphics[width=7cm,height=5.5cm]{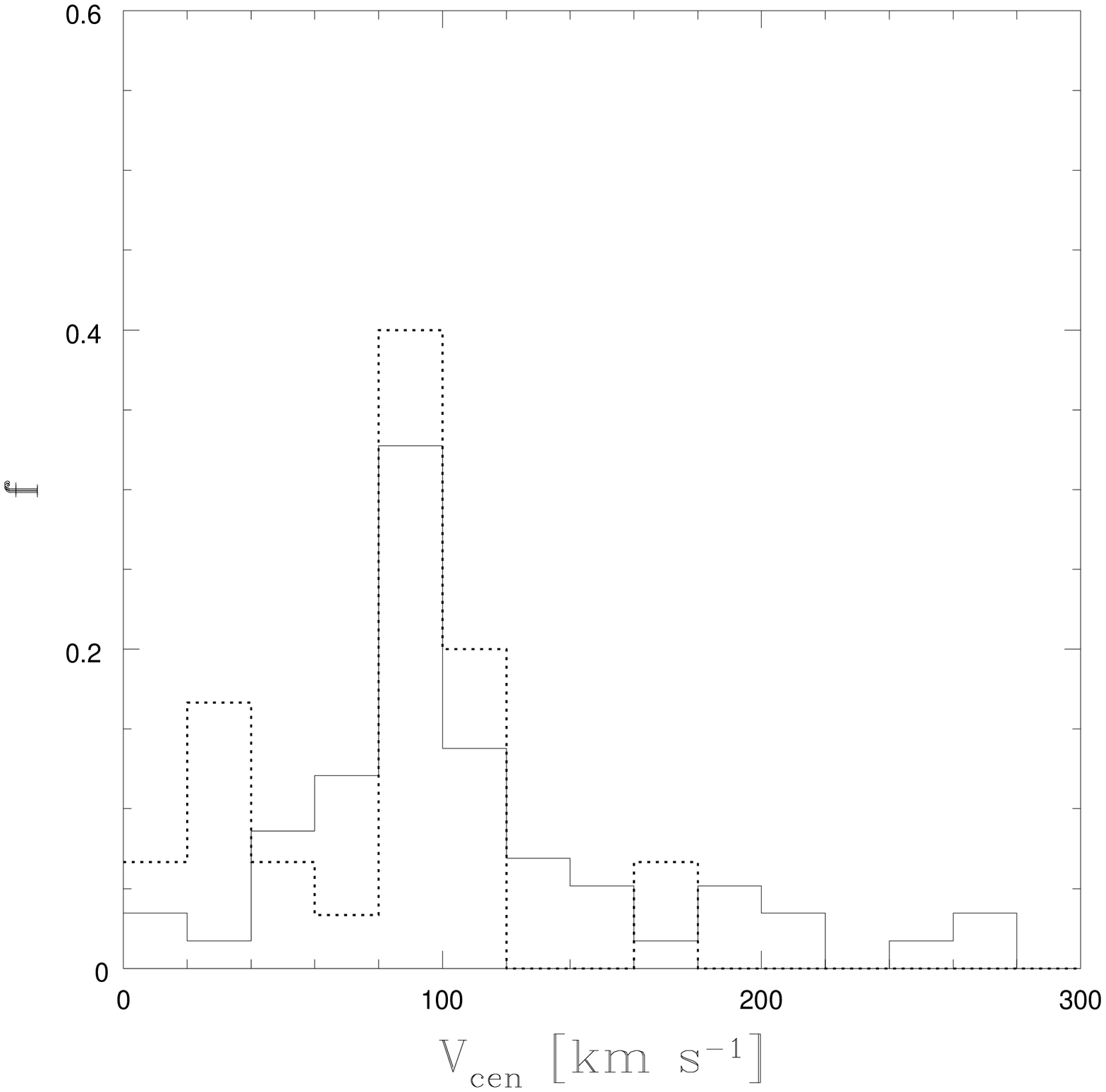}
\caption{Histogram of mean ages (upper panel) and central velocities (lower panel) for simulated galaxies in
pairs ($r < 100 \ {\rm km s^{-1}}$) separated according to their
star formation activity in the $\Lambda$CDM experiment.
 Dotted (solid) lines represent active (passive) star forming systems.
 }
\label{age}
\end{figure}

In Fig. ~\ref{age}a we show the distribution of mean stellar ages ($\tau$) for  active (dotted lines)
 and passive (solid lines) SF
simulated galaxies in pairs. As it can
be appreciated from this figure, passive SF galaxies in pairs show a peak distribution
with a mean age at $10.54$ Gyr $h^{-1}$. Systems undergoing strong SF activity show  a 
distribution shifted to shorter ages (with a mean at 7.65 Gyr $h^{-1}$) with
an important young tail. These results suggest that, on average, passive SF galaxies in 
pairs are dominated by an old stellar population.
Estimations of the gas fractions show  a trend for passive SF galaxies in pairs to have
smaller gas fractions with respect to active SF ones as it can be seen from Table 2
(see first two columns).

\begin{figure}
\centering
\includegraphics[width=7cm,height=5.5cm]{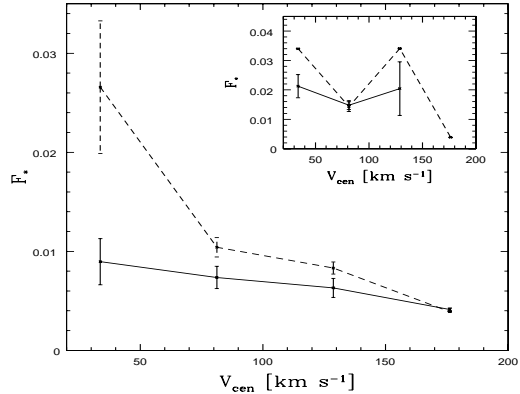}
\caption{Fraction of stars formed in the last $0.5$Gyr as a function of 
central circular velocity $V_{\rm cen}$ for passive star forming galaxies in pairs
with $r < 100 \ {\rm km \ s^{-1}}$ (solid lines) and in close pairs with $r < 30 \ {\rm km \  s^{-1}}$( dashed line)
in the $\Lambda$CDM experiment.
In the insert box we show the same distributions for active star forming galaxies in pairs.
 }
\label{fv}
\end{figure}

Another important parameter related to the capability of triggering star formation
by tidal torques is the characteristic of the potential well.
Previous works showed that  unless  dark matter haloes or compact stellar bulges 
 can provide stability to the systems, interactions
may induce strong gas inflows in gaseous discs triggering star formation activity (Athanassoula \& Sellwood 1986; Binney \& Tremaine 1987; Martinet 1995; 
Mo, Mao \& White 1998). Tissera et al. (2002) used the
central circular velocity ($V_{\rm cen}$) of the galactic systems to quantify the strength of the potential
well finding a correlation between the deepness of the potential well and the triggering
of the star formation activity. Following their work, we estimated the central circular velocity in
a similar fashion for galactic systems in pairs. As it can be seen from Fig.~\ref{age}b,  
we also found a trend for active SF systems to have less concentrated potential wells, with
a mean $V_{\rm cen}$ at 76 ${\rm km \ s^{-1}}$,
 than passive SF ones which
showed a mean $V_{\rm cen} $ at  103  ${\rm km \ s^{-1}}$. Note that the velocity distribution
of passive SF systems is quite broad and may include galactic objects with less concentrated
potential well which have just experienced a
 starburst.  We have calculated the fraction ($F_*$) of  total stellar mass which  formed in the last
0.5 Gyr  as a function of central circular velocity for these passive SF systems.
 We can see in Fig. ~\ref{fv} a clear correlation of the recent past
SF activity with the properties of the potential well (solid line). Systems with less concentrated
potential wells have  experienced a more important star formation activity in the recent past, although
all of them are currently forming stars at lower rates than the mean  of isolated galactic systems.
This trend is much stronger in close systems   ($r < r_{\rm c}$) where the combination of
proximity and shallow potential well correlates with  a more important SF activity in the recent past (dashed line).
We have also computed the corresponding fraction $F_*$ for active SF systems as depicted in  Fig. ~\ref{fv} (insert
box). As it can be seen from this figure, these systems also show a correlation of $F_*$ with $V_{\rm cen} $, although
they 
are experiencing a more important star formation activity with respect to their
past SF history than passive SF systems.  

In order to further investigate the effects of interactions on the SF activity, we
 assume that systems closer than $r_{\rm c}$ are merging candidates,
 while systems at larger
separation will be considered tidally interacting ones. 

\begin{itemize}

\item {\it Merging Pairs:}
We found that $66\%$ of simulated galaxies  in pairs are merging systems  and only  $38\%$ of them is actively forming stars.
 Passive SF merging pairs  have a mean gas fraction of $\approx 20\%$ while
active SF merging ones   have a  mean gas fraction of $30\%$. Passive SF systems
have older stellar populations and deeper potential wells than those with active
star formation as it can be seen from Table 2.

\item {\it Interacting Pairs: }
 We estimated that $10\%$ of the total  pair sample  are interacting systems with       enhanced
star formation activity, implying that $27\%$ of  pairs  with  $r >r_{\rm c}$   suffer
 tidally induced starbursts. Active interacting systems have larger gas fractions
and shallower  potential wells with
a mean at  $V_{\rm cen} $ at  62 \  ${\rm km s^{-1}}$. We also note that
passive interacting galaxies in pairs  have stars older than the rest of the pairs with a mean
age at 12.0 Gyr $h^{-1}$.  Table 2 summarizes these properties.

\end{itemize}

>From these results we  conclude that systems with shallower potential wells can
be affected by tidal interactions from larger distances than those with more concentrated
ones.   
Our findings also suggest that spatial proximity is
 a key factor in the strength  of the star formation
activity since the enhancement of SF detected for  close active  pairs doubled that
estimated for  active   interacting pairs  (Table 2).
We  found that in passive SF systems the combination of proximity and shallower potential
well can be related to an important star formation activity in the recent past
so that part of the currently passive star forming systems have experienced significant
SF in the last 0.5 Gyr.

Finally we can estimate the fraction of stars formed in galactic systems in pairs with respect
to the total stellar mass currently formed in the simulated volume. Recall that we are working
with a 10 ${\rm Mpc} \ h^{-1}$ box. In this simulated volume and considering systems with stellar
masses larger than $8 \times 10^{8} {M_{\odot} h^{-1}}$, we found that  pairs
closer than $r_{\rm c}$ contribute with  $34 \% $ of currently new born stars, while the total
pair sample is involved in the formation of  $42\%$ of the new stars at $z=0$.   

\begin{table*}
\begin{minipage}[t]{\columnwidth}
\caption{Mean physical properties of simulated galaxies in pairs in the 3D catalog.}
\label{table2}
\centering
\renewcommand{\footnoterule}{}  
\centering    
\begin{tabular}{lcccccccccc}
\hline
    & \multicolumn{2}{c}{Total Pair Sample
\footnote{Galaxy pairs with spatial separations $r < 100 \ {\rm kpc} \ h^{-1}$ }} &&
\multicolumn{2}{c}{Merging Pairs \footnote{Galaxy pairs with spatial separations $r < 30 \ {\rm kpc} \ h^{-1}$ }}
&&
\multicolumn{2}{c}{Interacting Pairs \footnote{Galaxy pairs with spatial separations $30 \ {\rm kpc} \ h^{-1} < r < 100 \ {\rm kpc \ h^{-1}}$ }  }
\\
  \cline{2-3} \cline{5-6} \cline{8-9}
   & Passive & Active && Passive & Active && Passive &  Active &\\
    \hline
Age\footnote{Mean stellar age given in Gyr $h^{-1}$}      &10.54 $\pm 0.01$ & 7.65$\pm 0.02$ && 9.25$\pm 0.02$ &  7.5 $\pm 0.05$&& 11.90 $ \pm 0.01$ & 7.80 $\pm 0.10$ \\
$\beta $ \footnote{Mean star formation enhancement} &0.36 $\pm 0.04$& 3.50$\pm 0.60$ &&  0.36 $\pm 0.05$ & 3.98 $\pm 0.07$ && $0.36 \pm 0.07$ & $2.09 \pm 0.29$\\
Gas\footnote{Percentage of the leftover gas in galactic systems}     &16  $\pm 7$ & 27 $\pm 7$ && 20 $\pm 3$ & $30 \pm 4 $ && $10 \pm 3$ &  $21\pm 4$\\
$V_{\rm cen}$\footnote{Central circular velocity in ${\rm km \ s^{-1}}$}  &103 $\pm 7$& 76 $\pm 7$ &&  120 $\pm 9$ & 90 $\pm 8$&& $86 \pm 11$ & $62 \pm 15$\\
\hline
\end{tabular}
\end{minipage}
\end{table*}

\subsection{Dependence on the Cosmological Parameters}

\begin{figure}
\centering
\includegraphics[width=7cm,height=5.5cm]{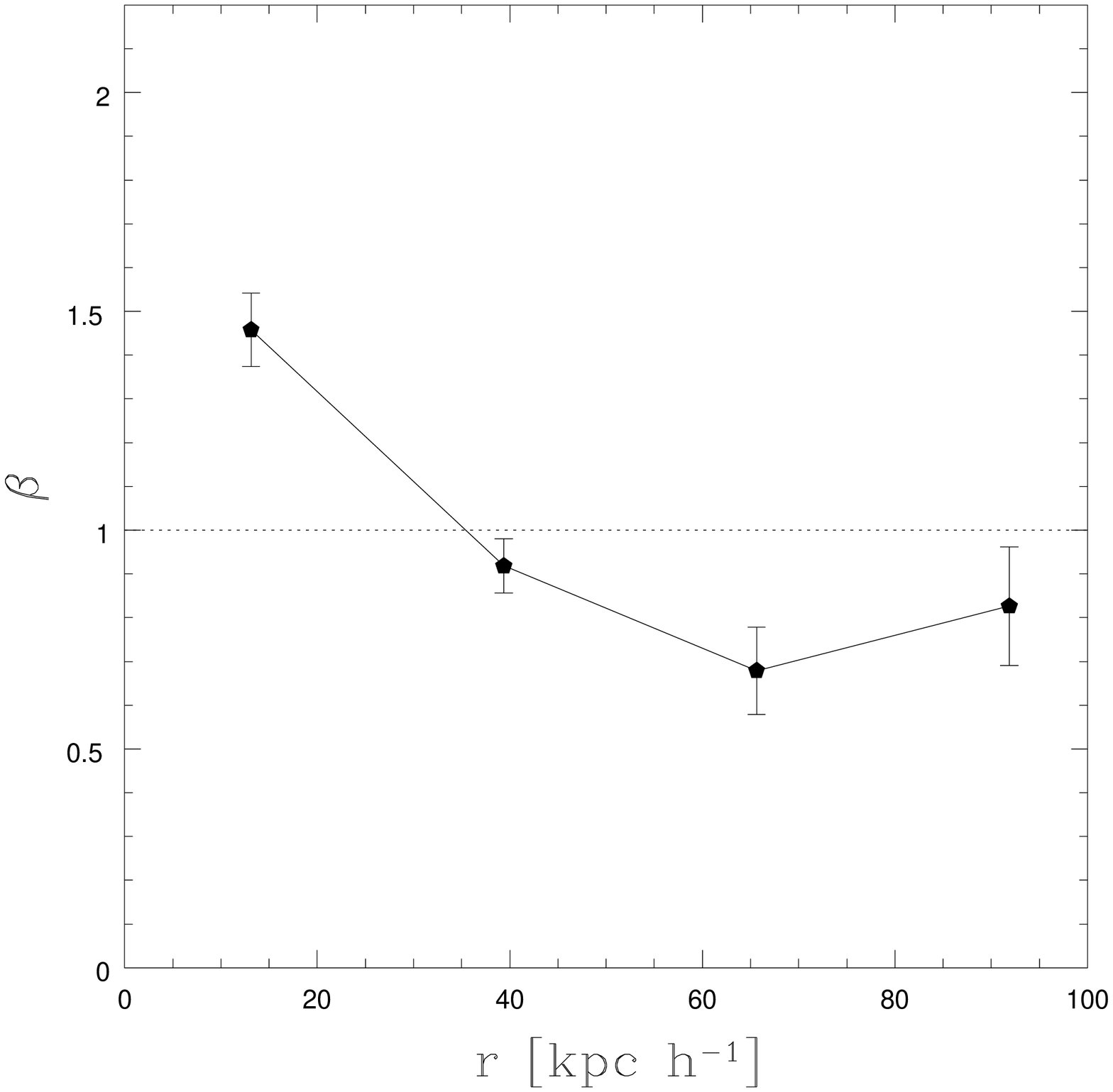}
\includegraphics[width=7cm,height=5.5cm]{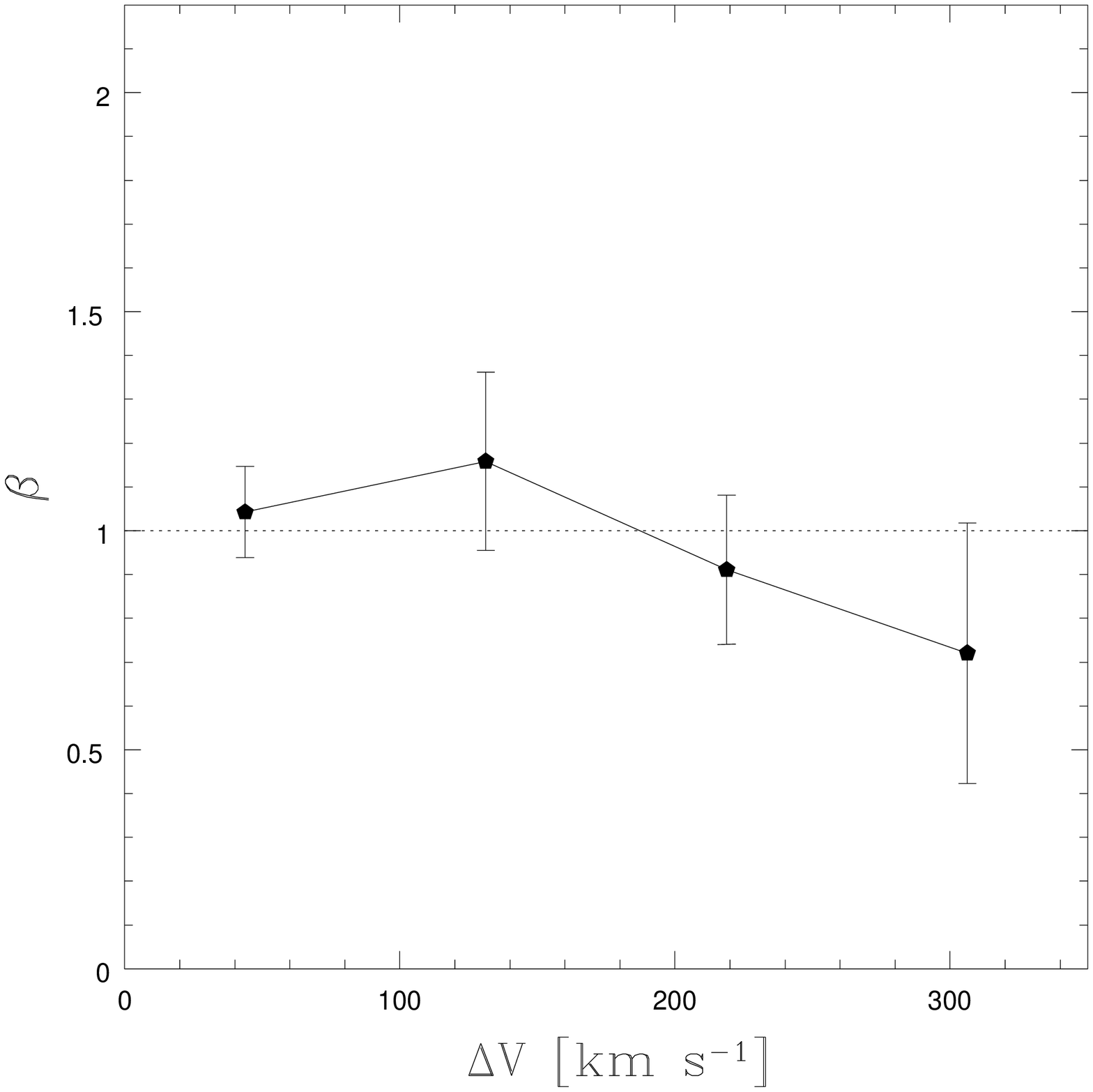}
\caption{Mean star formation excess parameter  $\beta $   
as a function of   tridimensional relative separation (upper  panel)
and velocity (lower panel) for galaxies in pairs 
in the $\Omega =1, \Lambda =0 $ CDM simulation.
}
\label{a3pm3d}
\end{figure}

Here we  discuss the results obtained from our low resolution SCDM run.
A similar analysis to that performed for  the 
$\Lambda$CDM simulation was carried out
in this run, building up a galaxy pair catalog from 
the 3D distribution of simulated galaxies. 
In  Fig.\ref{a3pm3d} we show the mean $\beta$ parameter 
as a function of distance and relative velocity
to the closest neighbour. The comparison of 
these trends with those found in  Fig.\ref{gad3d-d} and Fig. \ref{gad3d-v}
yields very similar results. In fact, 
the orbital parameter thresholds estimated for pairs in the SCDM simulation: 
$r_{\rm c} =  35\pm 5$ kpc $h^{-1}$  and  
$\Delta V_{\rm c} = 180\pm 30$ km s$^{-1}$, 
are in  good agreement with those
obtained for systems in the $\Lambda$CDM run. 

The general agreement at one $\sigma$-level found between the results from these two cosmological
scenarios probes  that the triggering of 
SF activity induced by galaxy-galaxy interaction 
is a local  physical mechanism which works 
independently of the cosmology (see also Section 5).

\section{Analysis of the 2D Galaxy Pair Catalog}

The star formation activity  in the  2D galaxy pair catalog 
in the $\Lambda$CDM run
 was analysed by  using the relative projected separations
 $r_{\rm p}$ and radial relative velocities $\Delta cz$. 

\begin{figure}
\centering
\includegraphics[width=7cm,height=5.5cm]{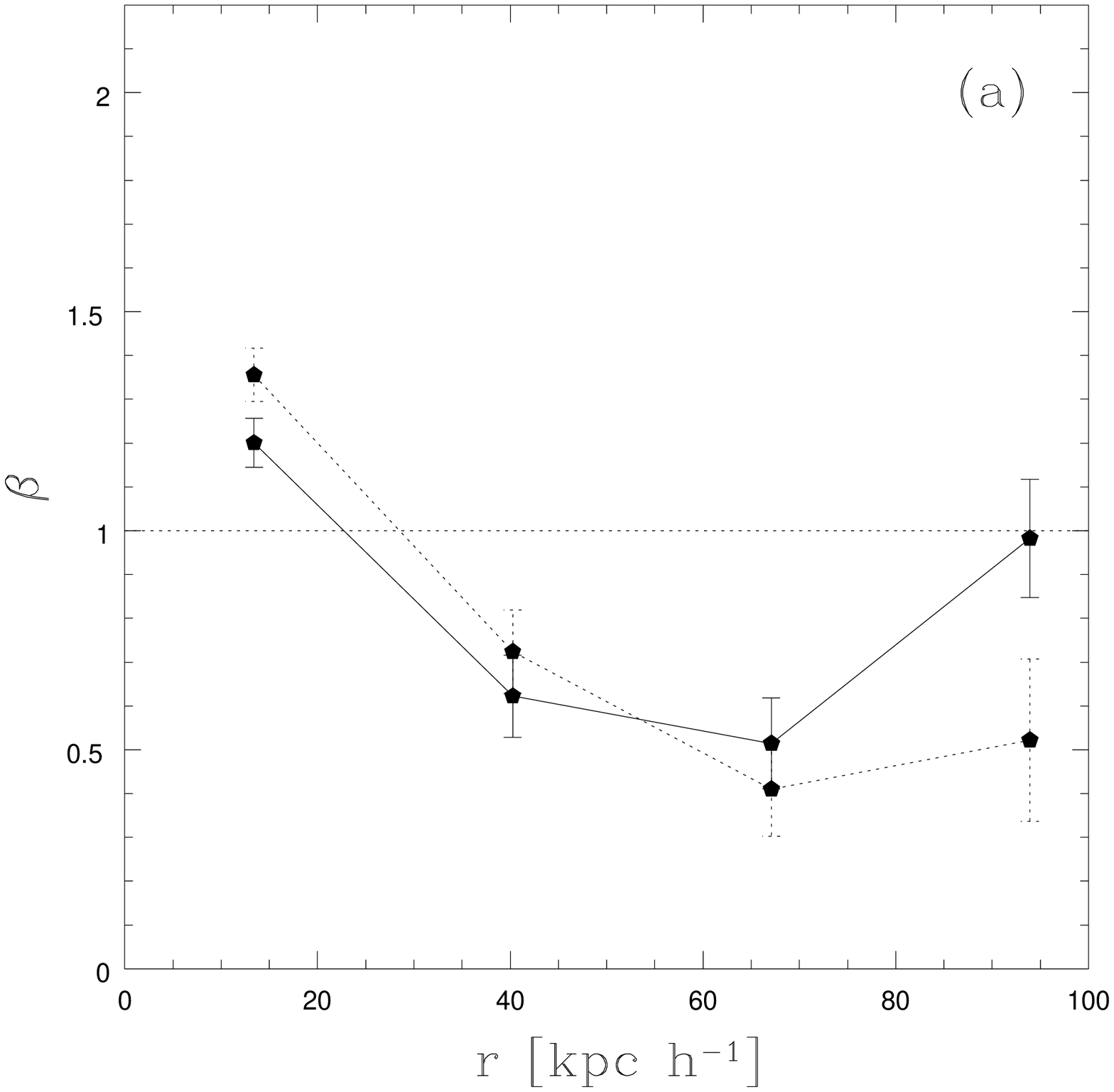}
\includegraphics[width=7cm,height=5.5cm]{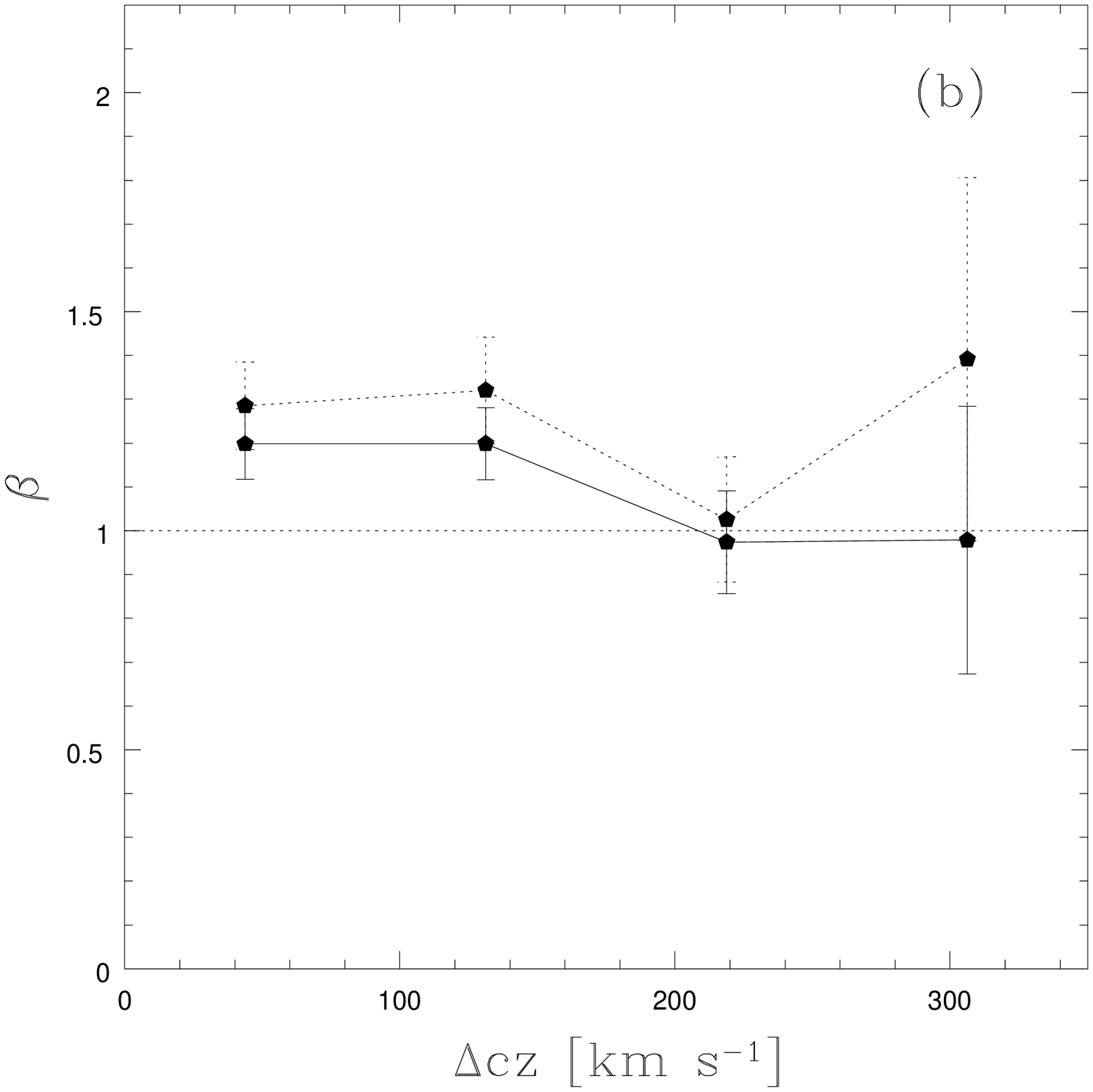}
\caption{Mean star formation excess parameter $\beta$  
as a function of   projected distance
 r$_{\rm p}$ (a) and $\Delta cz$ (b) for galaxies in 
pairs in the 2D simulated catalog (solid lines). 
In dotted lines are shown the   corresponding relations 
where spurious pairs have been removed } 
\label{spd-gad}
\end{figure}

As it
can be seen from Fig. \ref{spd-gad}a (solid lines), we found
a clear trend for an enhancement of the star formation activity  
with proximity in relative separation.
Comparing the trend in  Fig. \ref{spd-gad}a  
with that in Fig.  \ref{gad3d-d}, we can see that
the excess of star formation (i.e. $\beta > 1$ ) 
in the 2D-GP catalog  is detected for closer pairs  than in the case of the 3D-GP sample. 
This shrinking of the projected distance threshold 
compared to the 3D one  is produced by both 
 geometrical projection effects and spurious pairs.

In the case of  $\Delta cz$, we are  looking at 
the component of the velocity along
a random line-of-sight. This component
is strongly affected by the  geometrical orientation of 
the pair orbital plane  with respect to the chosen
random line-of-sight. Although this fact modifies   
the mean  star formation activity  signal, 
 the dependence on $\Delta cz$  is still  present as it can be
seen in Fig.  \ref{spd-gad}b (solid line).

>From this analysis we obtained that the  star formation  enhancement 
thresholds for the 2D-GP catalog are
$\bar{r_{\rm p}} = 25 \pm 5 \  {\rm kpc }\  h^{-1}$ and 
$\Delta \bar{cz} = 220 \pm 60 \ {\rm km \ s^{-1}}$. As we will discuss later on,
these values are in very good agreement with recent observational results (Lambas et al. 2003).

\subsection{Analysis of the effects of projecting galaxy pairs}

When projected data is analysed, as in the 2D-GP simulated catalog
or observations, some galaxies can appear as pairs when,
in fact, their tridimensional relative separation 
could be larger than the threshold values
adopted as a criteria to define pairs. 
 Hence, a study of these
effects in the simulation may help to understand 
their contribution to the observed trends.
 In this Section, we assess  the spurious effects 
introduced by both pairs in projection which have a tridimensional relative distance larger
than the adopted threshold of $100 \ {\rm kpc}\  h^{-1}$ (spurious pairs) 
and the distortion produced by the projection  of
3D pairs.

In the simulations,  spurious pairs are removed by checking if their tridimensional relative
distance is not within the adopted 3D threshold. 
 From this analysis we obtained that their effects are more
important at larger relative separations as expected.
The estimated percentages of spurious pairs
for the complete 2D-GP catalog  (i.e. $r_{\rm p}\leq 100$ kpc $h^{-1}$ and
$\Delta cz\leq 350$ km s$^{-1}$) is  $27\%$, 
while for  close 2D-GP sample (i.e. $r_{\rm p}\leq 25$ kpc $h^{-1}$ and
$\Delta cz\leq 100$ km s$^{-1}$) the contamination is smaller, $19\%$.

These results are in agreement with previous ones (e.g. Mammon  1986; Nikolic et al. 2004).  
In particular, Alonso et al. (2004) estimated the percentage of spurious pairs
by using the  2dFGRS mock catalog constructed by  Mech\'an \& Zandivarez (2002),
finding that for the complete mock pair catalog, $29\%$ were
spurious pairs while for the close pair one, the percentage was smaller ($21\%$).

In Fig. \ref{spd-gad}, we  show the dependence of the SF activity
on $r_{\rm p}$ and $\Delta cz$ for 
 the  2D-GPs catalog (solid line) and for the 2D-GPs with spurious pairs
being removed (dotted line). The mixing of arbitrary values of SF 
by spurious pairs is expected to produce a
 lowering of both  the  star formation activity signal  
and   the  relative distance threshold
$\bar{r_{p}}$.  These effects can be appreciated 
in Fig. \ref{spd-gad}a from where we can see that, 
when spurious pairs are eliminated (dotted lines), 
an  increase of $ 30 \% $ and $10\%$ in  the relative distance threshold and
 the star formation enhancement signals, respectively, are detected.
 On the other hand, by  comparing  the  trends 
of the  3D-GP catalog (Fig. ~\ref{gad3d-d})  and 2D-GP one (Fig. ~\ref{spd-gad}) with spurious pairs removed,  we found that the effects 
of projection  produced 
a decrease of  the signal  of  star formation enhancement 
by $\approx 25\%$ and of the relative separation threshold by  less than $\approx 10\%$.  

Our results suggest that the projection of 3D pairs seems to affect more strongly 
the star formation activity signal than  the $r_{\rm p}$ threshold, while
spurious pairs seem to have a more important impact on  the shrinking of the $r_{\rm p}$ threshold.

\subsection{Dependence of Star Formation on Environment}  

In this section, we focus on the effects of galaxy interactions on SF 
activity by taking into account their local environment in order to analyse 
if $\Lambda$CDM cosmologies  can reproduce the observational 
trends detected in different works (Gomez et al. 2003;
Balogh et al. 2004; Alonso et al. 2004).
For this purpose we use the simulated 2D-GP catalog discussed in previous sections
and the corresponding 3D-GP sample.

In order to  characterize the local environment of   simulated galaxies 
in a similar way to that employed  in observations (Dressler et al. 1980;
Gomez et al. 2003; Balogh et al. 2004), 
we  calculated  the projected local density parameter
 as $\Sigma=6/(\pi d^{2})$ where {\it d} is  the projected distance 
to the 6$^{th}$ neighbour brighter than M$_{\rm r} = -20.5 $ (we assumed a mass-to-light ratio of 5 to convert
stellar masses to luminosities) and $\Delta cz<1000$ km $s^{-1}$ (Balogh et al. 2004).

\begin{figure}
\centering
\includegraphics[width=7cm,height=5.5cm]{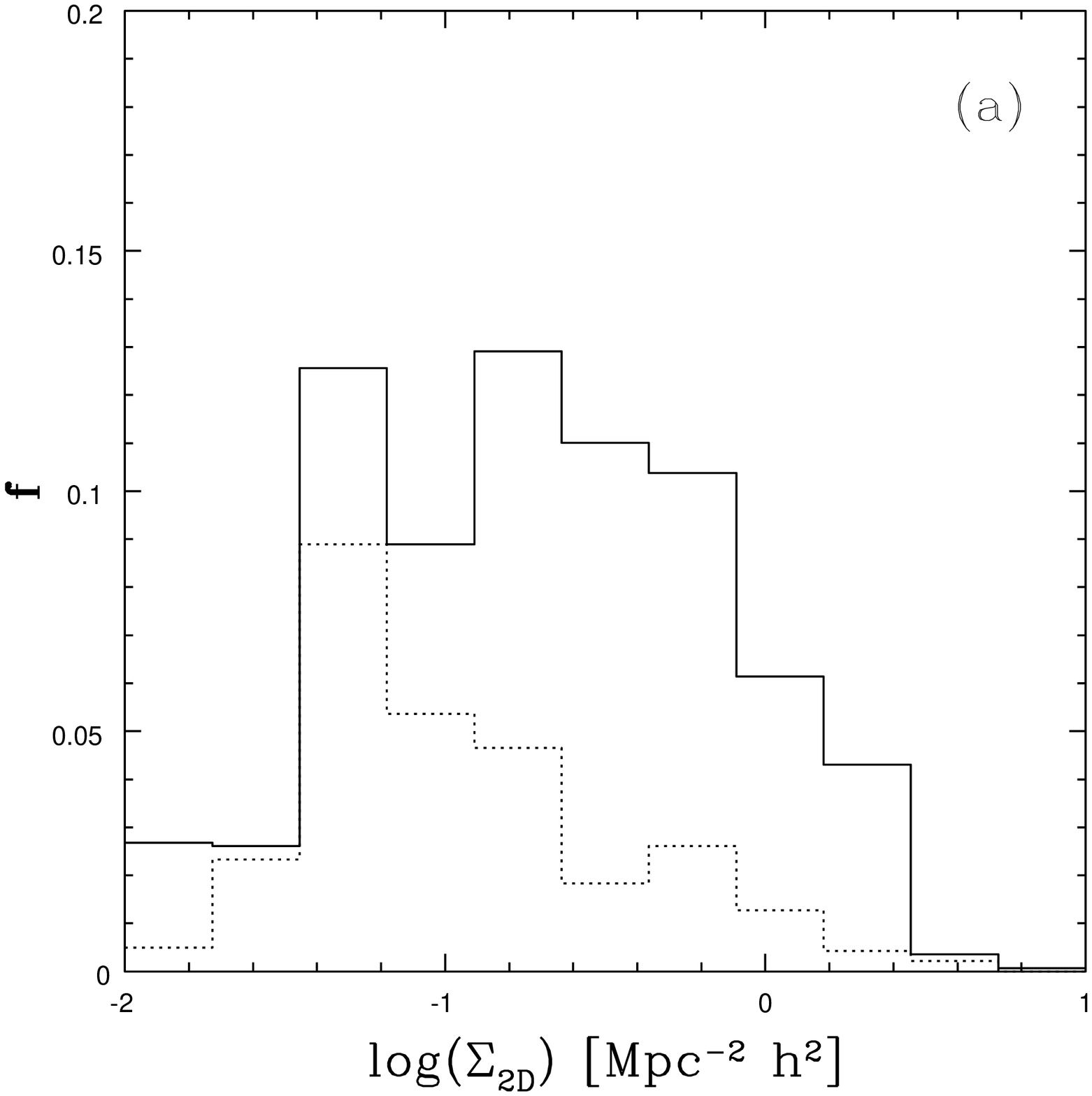}
\includegraphics[width=7cm,height=5.5cm]{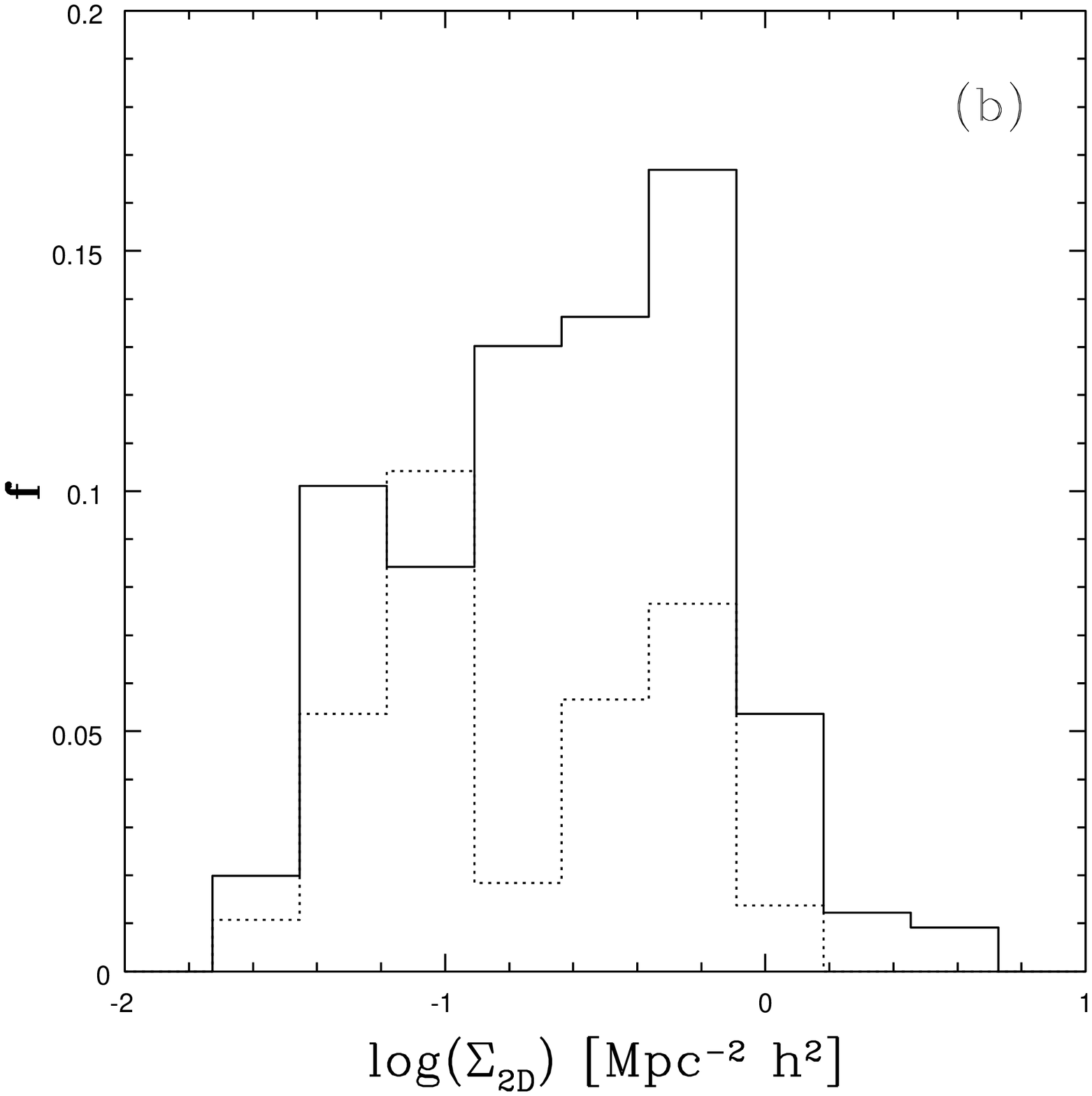}
\caption{Distribution of the fraction of active (dotted line) and
passive  (solid line) SF galactic systems  
in the projected  control (a) and pair (b) catalogs
as a function of projected density estimator $log(\Sigma_{\rm 2D})$.
}
\label{histo-sig}
\end{figure}

In Fig.\ref{histo-sig}, we show the distributions of galaxies according to their local density $\Sigma$ 
for simulated galaxies without a near companion (a) 
and simulated galaxies in pairs (b),
 separated in two groups depending on their
star formation activity:   passive star-forming  galaxies   
(solid lines) and active star-forming galaxy 
 (dotted lines), as defined in Section 3. 
As we can see from Fig.\ref{histo-sig}a, 
passive star forming simulated galaxies 
without a close companion  dominate the   higher density regions, 
while active SF ones  prefer the lower density environments\footnote{A stronger SF-density
relation is obtained from the simulations if a more restrictive value of $\beta$ is used
to classify systems are passive or active. Note that $\beta =1$ implies that a system is forming
stars at a rate comparable to the average of the Milky Way.}, accordingly to
the observed SFR-density relation (Gomez et al. 2003) 
and its analogous density-morphology relation (Dressler 1980). A similar trend is found
for galaxies in pairs as it can be seen in Fig. ~\ref{histo-sig}b.

In order to quantify this trend, we adopt a value of $log\Sigma =-0.80$ 
to separate galaxies into  low and high density regions.
 This $log\Sigma=-0.80$
value segregates the sample in two subsamples with similar number of members.
>From Table 3 we can see that, the ratio ($ P/A$)  between the percentage of passive and
active SF galaxies decreases from 3.1 in high density regions to 1.6 in low density ones for the control sample. 
 A similar behaviour is found for simulated galaxies in pairs with  $ P/A $ varying from $ \approx 2.5$
to  $  \approx 1.4$ from high to low density regions, respectively.
We found  no significant differences between the dependence of SF on environment for galaxies with or without
a companion. 
However, if the systems are in   close pairs ($r_{\rm p}\leq 25$ kpc $h^{-1}$ and $\Delta cz \leq$ 100 km s$^{-1}$), 
 the dependence of $ P/A$ on environment is found to
be stronger, changing from  $ P/A \approx 0.7$ in low density regions to $ P/A \approx 2.6$  in
high density ones.

\begin{figure}
\centering
\includegraphics[width=6cm,height=5.5cm]{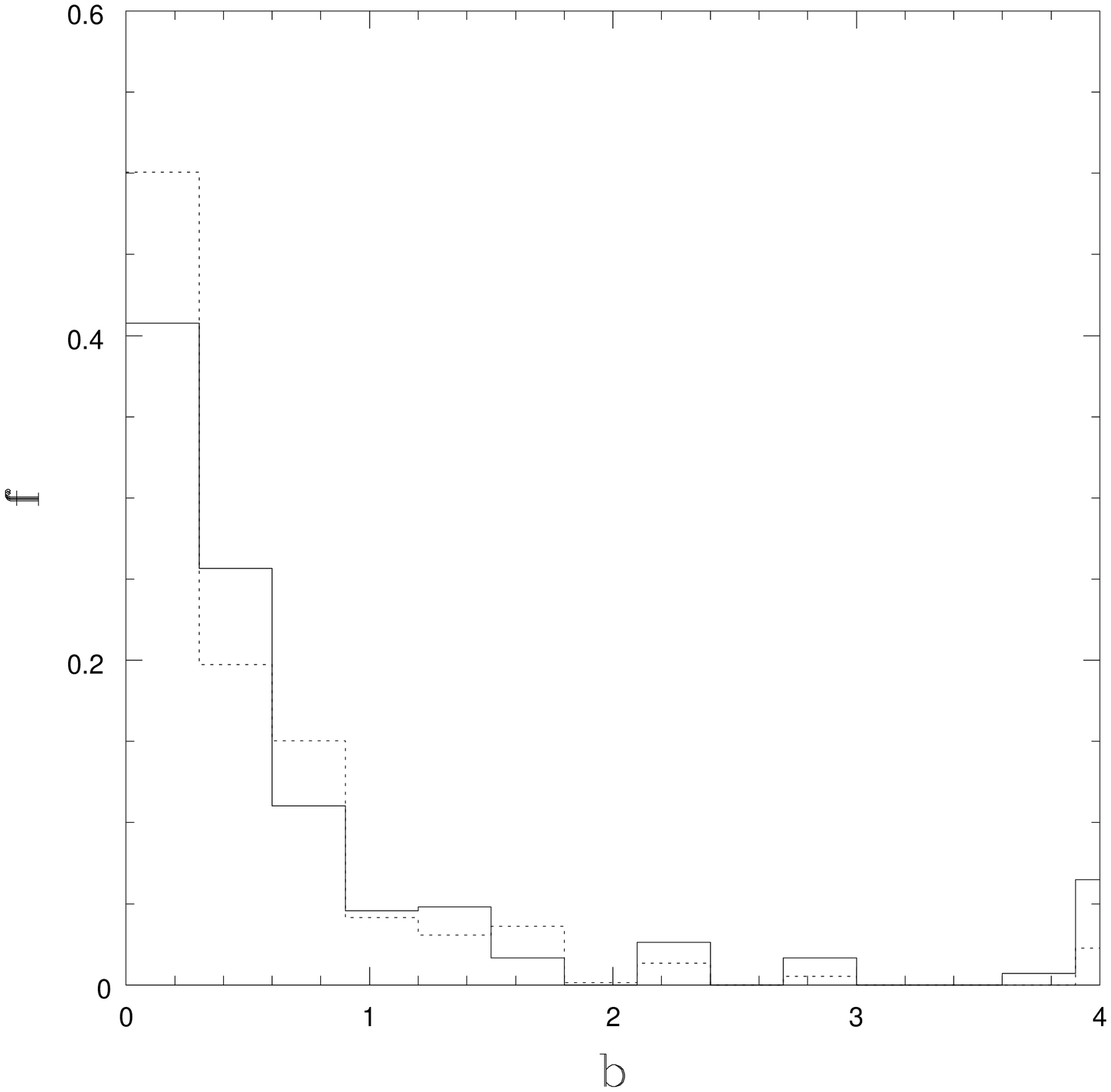}\\
\includegraphics[width=6cm,height=5.5cm]{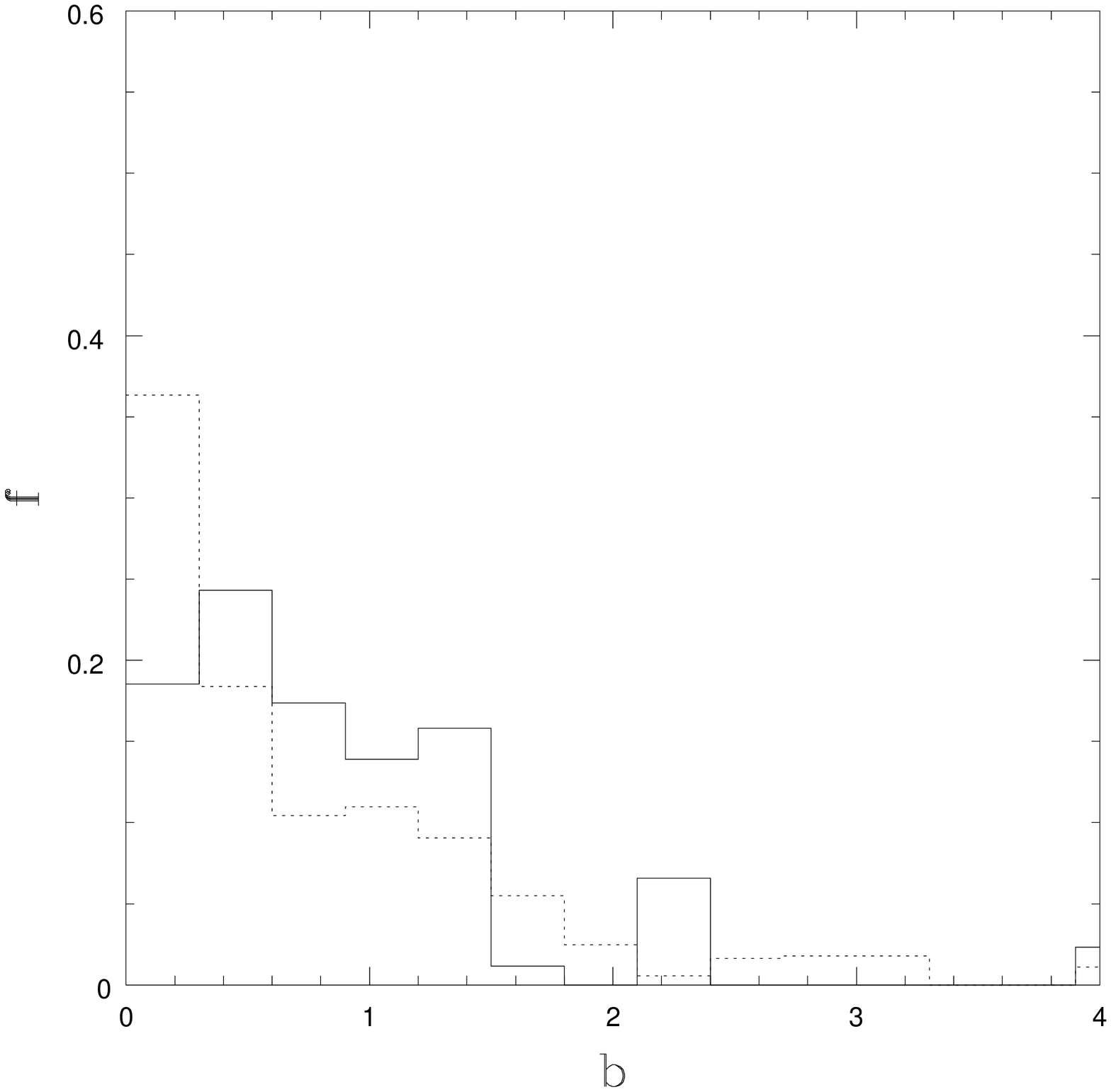}
\caption{Histograms of the birth rate parameter (b) for galaxies in pairs (solid) and
galaxies without a near companion (dashed), in high (upper panel) and low (lower panel) {\bf density}
regions.
}
\label{sfr-den}
\end{figure}

In  Fig.~\ref{sfr-den} we show the distribution of birthrate parameters $b$ for galaxies in pairs
in low (lower panel) and high (upper panel) density regions and the distributions for their corresponding control samples.
As it can be appreciated from this figure the SF activity tends to be larger in galaxies with a close companion,
principally in low density regions. 
We computed  the median ($b_{\rm m}$) of these distributions and  found that for the control and galaxy pair samples,
the $b_{\rm m}$ values increase from high to low density regions by a similar factor (1.6 and 1.7, respectively).
Conversely,  close pairs show  a stronger dependence of this parameter on environment by changing a factor of 3
in the same density range. The $b_{\rm m}$ parameter of close galaxy pairs in high density regions
is comparable to that of the control sample ($b_{\rm m}\approx 0.37$) while in  low density environments,
 the $b_{\rm m}$ parameter of close pairs exceeds by  a factor of three that of the control sample.

The important decrease in
the star formation activity and the significant increase of the fraction of passive SF members from low to high 
density regions for galaxies in close pairs suggest that interactions play a relevant role in the origin of the 
SF-density relation.
As shown in Section 3,  passive star forming systems
tend to be gas poor, to be dominated by old stellar populations and to have deep potential wells,
 indicating that they are in an advanced stage of evolution. In this sense the fact
that in high density regions a larger fraction of galaxies in close  pairs are passively forming star
shows that the environment plays a role by accelerating the structure evolution 
as expected in hierarchical clustering scenarios where mergers and interactions are
more common in high density environments.

So far we have not taken into account the effects of spurious pairs. 
We estimated that for the projected simulated-galaxy pairs
in low density environments  (i.e. $r_{\rm p}\leq 100$ kpc $h^{-1}$ and
$\Delta cz\leq 350$ km s$^{-1}$), $17\%$   are spurious  pairs
while this percentage increases to $33 \%$ in high density regions. 
For  close pairs ($r_{\rm p}\leq 25$ kpc $h^{-1}$ and $\Delta cz \leq$ 100 km s$^{-1}$), these
percentages go down to $14\%$ and  $23\%$ for low and high environments, respectively.
If the 2G-PG catalog is cleaned up of these spurious pairs, 
the dependence  of the star formation activity on environment remains basically
 unchanged.
For this clean sample, we found  a $P/A$ ratio of $\approx 2.56$ and $\approx 1.26$ in high and low density regions, respectively.
These numbers are comparable to those found for the complete 2G-PG catalog (Table 3).

\begin{figure}
\centering
\includegraphics[width=6cm,height=5.5cm]{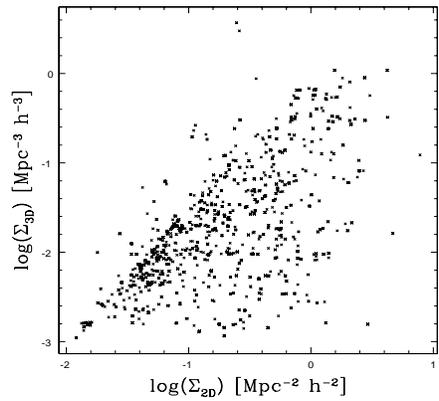}
\caption{Correlation between the density estimator $\Sigma$ calculated by using the distance to 
the $6^{th}$ neighbour in the 3D galaxy distribution  and in the projected one.}
\label{sigma2d3d}
\end{figure}

We have also calculated a  density estimator in three dimensions
as $\Sigma \propto 6/(\pi d^{3})$ where {\it d} is now  the tridimensional distance 
to the 6$^{th}$ neighbour brighter than M$_{\rm r} = -20.5 $. As it can
be seen from Fig.~\ref{sigma2d3d} the  3D local densities (i.e. $\Sigma_{\rm 3D}$)  of these galaxies are
lower than that estimated from the projected distances ($\Sigma_{\rm 2D}$). This suggests that  projected density
estimators  can   overpredict the local density.
Note however that there is a  correlation between both estimators although with an  important dispersion towards
larger values of $\Sigma_{\rm 2D}$.
 It is worthnoting that our simulation corresponds to
a typical field region, so that most  galaxies  have low   $\Sigma_{\rm 3D}$ with a minimum possible value
at log $\Sigma_{\rm 3D}= -3.60 $ defined by the size of the box. For the projected sample this minimun  value
increases to log $\Sigma_{\rm 2D}= -2.20$ due to projection effects.

\begin{figure}
\centering
\includegraphics[width=7cm,height=5.5cm]{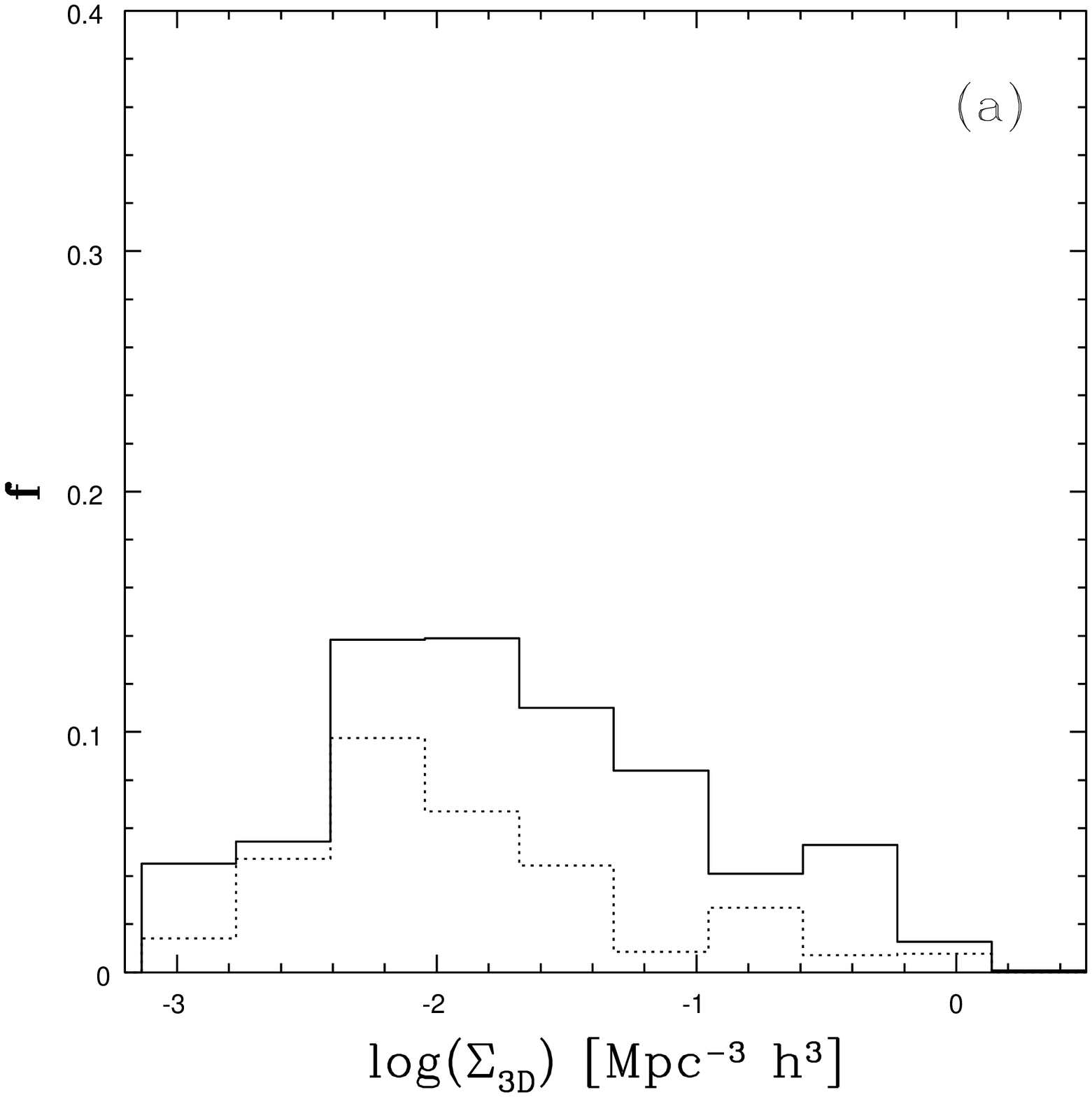}
\includegraphics[width=7cm,height=5.5cm]{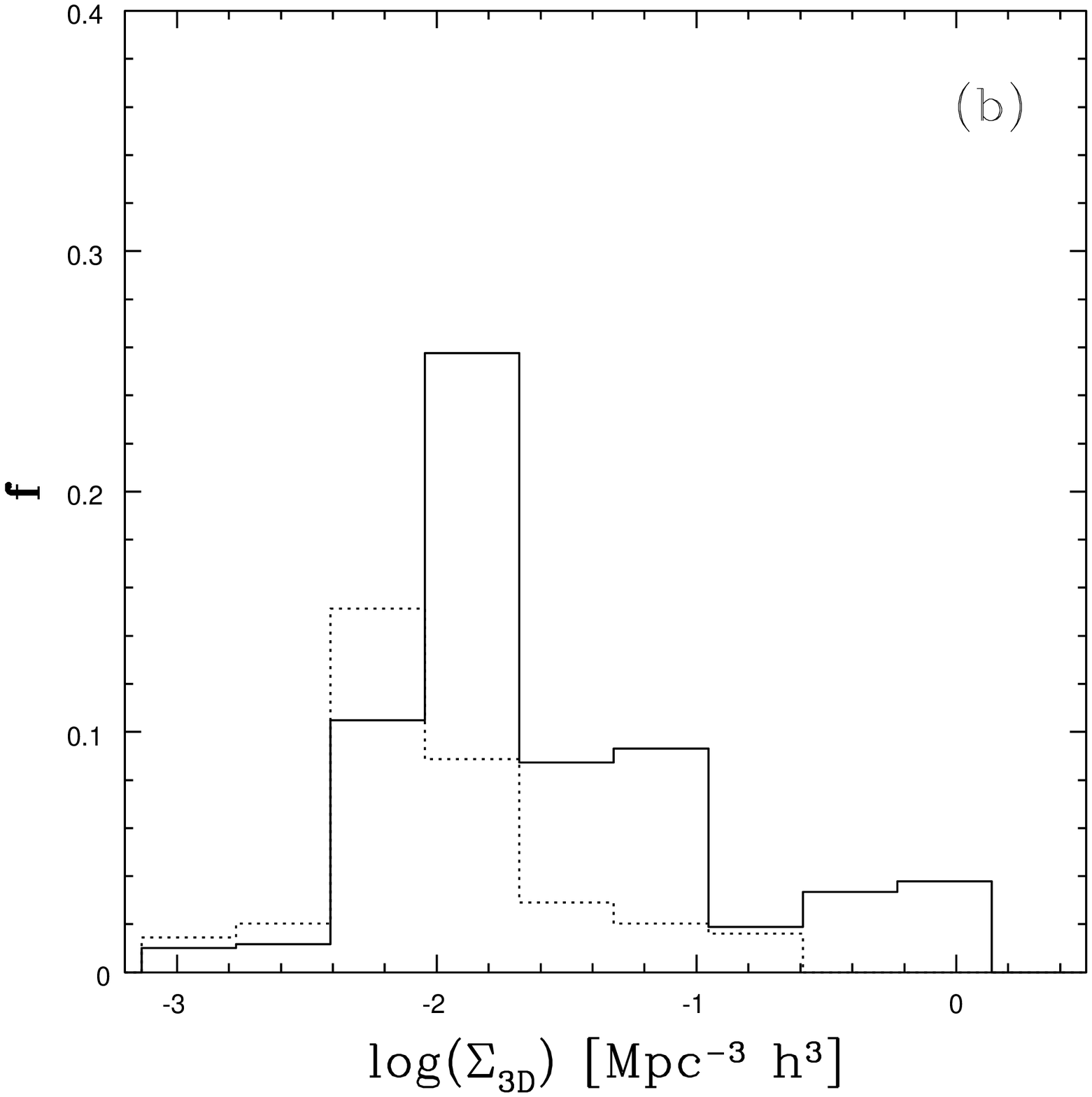}
\caption{Distribution of the fraction of galaxy-like systems  
in  control (a) and pairs (b) catalogs
as a function of $log(\Sigma_{\rm 3D})$ estimated from the 3D distance to the neighbours.
Simulated galaxies have been divided in two groups:
active  (dotted line ) and 
passive  (solid line) star forming sytems. 
}
\label{histo3d}
\end{figure}

In Fig.~\ref{histo3d} we show similar distributions to those of Fig.~\ref{histo-sig} but  as a function 
of  $\Sigma_{\rm 3D}$. We have also estimated the fractions $P/A$ for these distributions  taking
the value   $\Sigma_{\rm 3D}= -1.75 $ to divide the samples into  low and  high density subsamples with comparable number of members.
The $P/A$ fractions for these samples indicate  a weaker SF-density relation for the control sample, 
with a variation of  $P/A$ from $\approx 2.8$ to $\approx 1.8$ from high to low density regions, and a stronger
one for galaxies in pairs, with  $P/A$ varying from $\approx 3.0$ to $\approx 1.4$ in the same density range.
Hence, overall, projection seems to weaken the role of pairs in the SF-density relation.

\begin{table*}
\begin{minipage}[t]{\columnwidth}
\caption{Dependence of SF activity on the environment for the projected 2D catalogs.}
\label{table2}
\centering
\renewcommand{\footnoterule}{}  
\centering    
\begin{tabular}{lccccccccccccc}
\hline
    & \multicolumn{3}{c}{Total Pair Sample
\footnote{Galaxy pairs with spatial separations $r_{\rm p} < 100 \ {\rm kpc} \ h^{-1}$ and $\Delta cz < 350 \ { \rm km \ s^{-1}}$}} &&
\multicolumn{3}{c}{Close Pairs Sample\footnote{Galaxy pairs with spatial separations $r_{\rm p} < 25 \ {\rm kpc} \ h^{-1}$ and $\Delta cz < 100 \ { \rm km \ s^{-1}}$ }}
&&
\multicolumn{3}{c}{Control Sample \footnote{Galaxy pairs with spatial separations $r_{\rm p} > 100 \ {\rm kpc} \ h^{-1}$ }  }
\\
  \cline{2-4} \cline{6-8} \cline{10-12}
   & Passive \footnote{Passive and Active columns represent percentages normalized to the total number of members in each Total Sample. We use log $\sigma =-0.8$ to define high/density environments}  & Active&$P/A$ \footnote{Ratio between the percentage of passive ($P$) and active ($A$) galaxies.} && Passive & Active & $P/A$&& Passive &  Active&$P/A$ &\\
    \hline
Total Samples     &66 & 34&--&& 55 &  45&--&& 68 & 32 &--\\
High density regions &43& 17 &2.53&&  37 & 18 &2.06&& 37 & 12&3.08\\
Low density regions     &23 & 17 &1.35&& 18  & 27 &0.67&& 31 & 20&1.55  \\
\hline
\end{tabular}
\end{minipage}
\end{table*}


\subsection{Comparison with observations}

The results discussed in previous sections  lead us to conclude that
even  if in 2D we only considered  tridimensional pairs, projection redistributes
their contribution  in  projected distance and 
relative velocity bins.

\begin{figure}
\centering
\includegraphics[width=7cm,height=5.5cm]{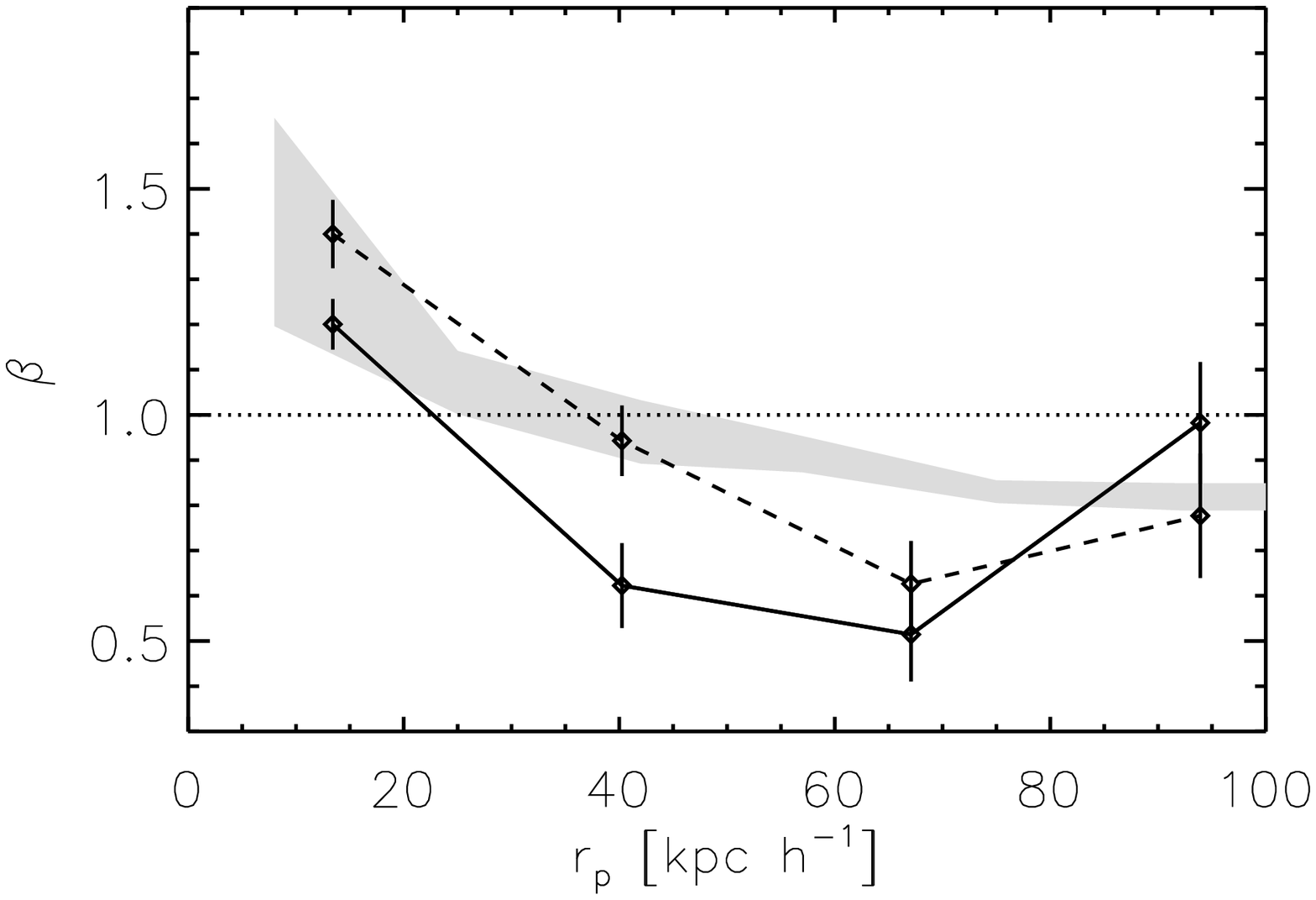}
\includegraphics[width=7cm,height=5.5cm]{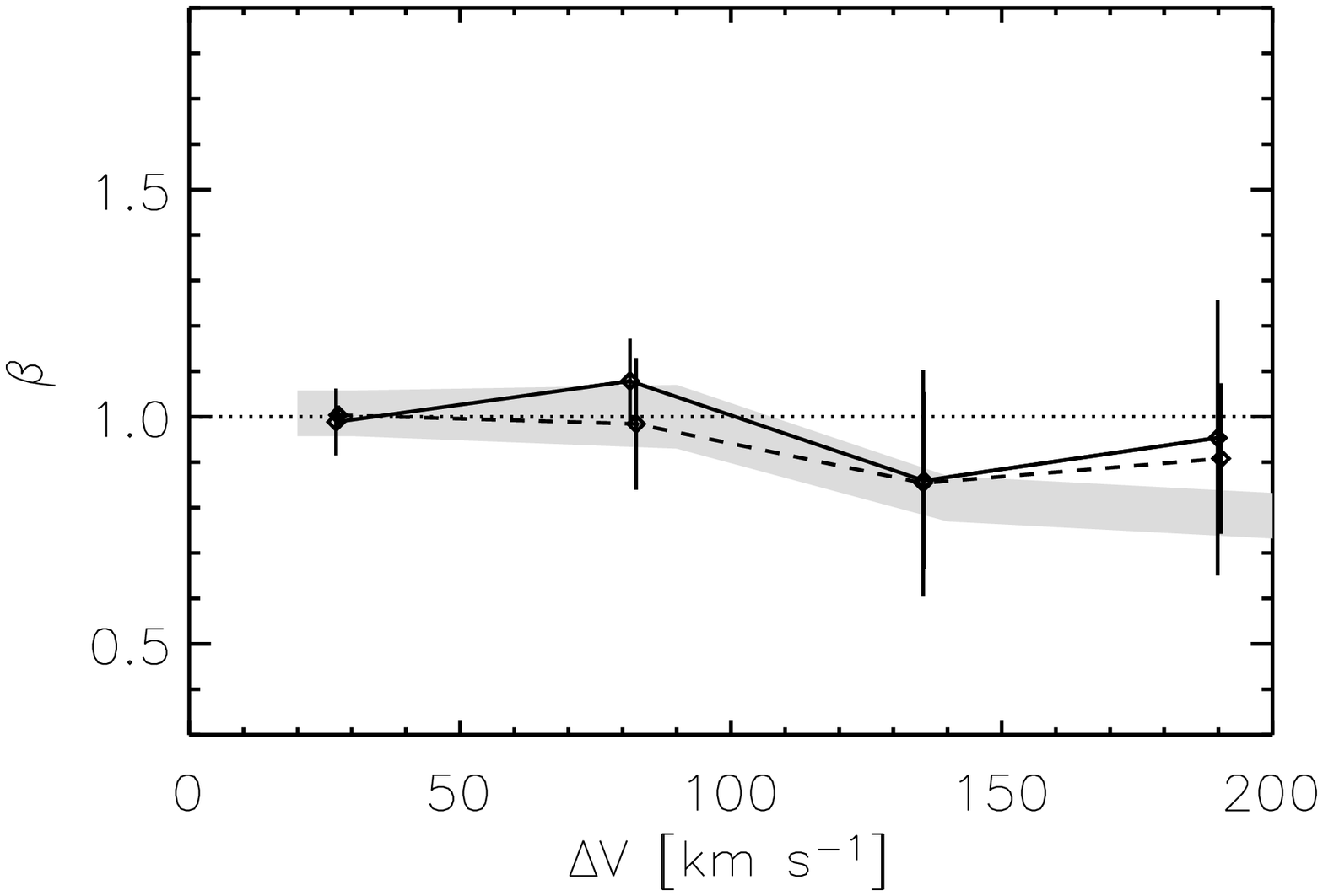}
\caption{Mean star formation excess parameter $\beta$ 
as a function of   projected distance
 r$_{\rm p}$ (upper panel) and radial relative velocities (lower panel) for galaxies in 
pairs of the 2D catalog in the $\Lambda$CDM ($\Omega=1,\, \Lambda=0.7 $) 
(solid line) and SCDM ($\Omega=1,\, \Lambda=0 $) (dashed line).
The shades areas represent 
to the observed relations obtained by Lambas et al. (2003)  
for galaxy pairs in the field in the 2dFGRS. The area width is determined
by the corresponding observed dispersion. } 
\label{comp-r}
\end{figure}

In Fig.~\ref{comp-r} we show the 2D-GPs dependence of 
star formation on $r_{\rm p}$ and $\Delta cz$ for the $\Lambda$CDM (solid line)
and SCDM (dashed line),
including the observational trends obtained by Lambas et al. (2003) 
 for galaxy pairs in the field. In the case of 
the projected radial velocity, we introduced a random noise in the simulated relations with
maximum amplitude equal to two times the observational velocity error 
of the 2dFGRS (Coles et al. 2001)
 in order to make the comparison with observations more realistic. 
 As it can be appreciated, the relations are 
within the observed trends, although in the case
of relative separation  the simulated star formation activity at larger distances 
is lower than the observed one. Nevertheless, 
we note that observations and simulations
are in agreement at 1.5 $\sigma$ level. 

Since for the simulated pairs the enhancement of 
star formation during the interactions are
produced by tidal torques, the agreement found with observations suggests that 
the observed correlations are produced by  this mechanism 
and seems to be  independent of cosmology.

\section{Conclusions}

In this paper we analysed the star formation activity 
in galactic systems in pairs in a $\Lambda$CDM and in a SCDM
scenarios. We also assessed effects introduced when  
pairs are selected from a 2D projection of
the tridimensional galaxy  distribution.

Our findings can be summarized as follows:

\begin{enumerate}
\item  We found that tidal interactions trigger star formation 
at levels higher than those statistically measured for 
galactic systems without a close companion, if systems
are closer than 30 kpc $h^{-1}$. We  did not find a clear difference 
between low and high velocity encounters.
By comparing with results obtained with a different cosmological model, we
infer that interaction-induced star formation is a generic feature whose
trends do not depend strongly on cosmology. 

\item Our findings showed that dynamically unstable systems with shallow potential wells
 could experience 
gas inflows driven by tidal torques at earlier stages in the interaction. 
Conversely, stable systems need to have a closer neighbour to experience strong star formation activity.
These results suggest that the combination of spatial proximity and the characteristics
of the potential wells  seems to be a key factor to induce strong star formation
activity during an interaction.

\item We also found that close systems with no SF enhancement  have, 
on average,  steep potential wells, low gas fraction and are dominated by  old stars. However,
some of them have shallow mass distributions and have experienced some SF activity
in the near past, so that the fraction of recently formed stars anticorrelates 
with  central circular velocity.  These systems
are, on average, in a more advance stage of evolution in comparison
with galaxies in pairs with strong star formation activity.

\item  The projected simulated galaxy pair catalog 
showed star formation activity in very good agreement with
observations. We detected that, in projected distance, 
systems have to be closer than $ \approx 25 $ kpc $h^{-1}$
to show important star formation enhancement in agreement with observations.
\item  We found that spurious pairs represent 27 $\%$ 
of the 2D-GP sample but this percentage decreases
to $19 \%$ for close pairs. Spurious pairs are more likely 
to affect high density regions ($33\% $
of spurious pairs are detected in the high density regions 
while $17\%$ corresponds to the low
density environments).

\item  The analysis of the dependence of the star formation 
on local density yields a relation in agreement
with the observed SF-density one for both galaxies in pairs and in the control sample, 
with a comparable decrease in the level of SF activity and a comparable  increase in the  fraction
of passive SF systems with increasing local density.
However, these trends are significantly stronger for close pairs which 
 suggests that, in a hierarchical clustering
scenario, galaxy-galaxy interactions contribute to establish a   SF-density relation. 

\end{enumerate}

\begin{acknowledgements}
We are grateful to the anonymous referee for the  thorough remarks and comments which 
importantly helped to improve this paper.
The simulations were run on the Ingeld PC-cluster funded by Fundaci\'on Antorchas.
This work was partially supported by the
Consejo Nacional de Investigaciones Cient\'{\i}ficas y 
T\'ecnicas and Fundaci\'on Antorchas. 
\end{acknowledgements}

\end{document}